\documentclass[twocolumn,showpacs,preprintnumbers,amsmath,amssymb]{revtex4}

\usepackage{graphicx}
\usepackage{dcolumn}
\usepackage{bm}

\newcommand{\bq}{\begin{equation}}
\newcommand{\eq}{\end{equation}}
\newcommand{\bqa}{\begin{eqnarray}}
\newcommand{\eqa}{\end{eqnarray}}
\newcommand{\nn}{\nonumber \\}

\def\be     {\begin{equation}}
\def\ee     {\end{equation}}
\def\bea        {\begin{eqnarray}}
\def\eea        {\end{eqnarray}}
\def\bnn    {\begin{eqnarray*}}
\def\enn    {\end{eqnarray*}}

\begin{document}

\title{Superconductivity from a non-Fermi liquid metal :
Kondo fluctuation mechanism in the slave-fermion theory}
\author{Ki-Seok Kim}

\affiliation{Asia Pacific Center for Theoretical Physics, Hogil
Kim Memorial building 5th floor, POSTECH, Hyoja-dong, Namgu,
Pohang 790-784, Korea}

\date{\today}

\begin{abstract}
We propose Kondo fluctuation mechanism of superconductivity,
differentiated from the spin fluctuation theory as the standard
model for unconventional superconductivity in the weak coupling
approach. Based on the U(1) slave-fermion representation of an
effective Anderson lattice model, where localized spins are
described by the Schwinger boson theory and hybridization or Kondo
fluctuations weaken antiferromagnetic correlations of localized
spins, we found an antiferromagnetic quantum critical point from
an antiferromagnetic metal to a heavy fermion metal in our recent
study. The Kondo induced antiferromagnetic quantum critical point
was shown to be described by both conduction electrons and
fermionic holons interacting with critical spin fluctuations given
by deconfined bosonic spinons with a spin quantum number $1/2$.
Surprisingly, such critical modes turned out to be described by
the dynamical exponent $z = 3$, giving rise to the well known
non-Fermi liquid physics such as the divergent Gr\"{u}neisen ratio
with an exponent $2/3$ and temperature-linear resistivity in three
dimensions. We find that the $z = 3$ antiferromagnetic quantum
critical point becomes unstable against superconductivity, where
critical spinon excitations give rise to pairing correlations
between conduction electrons and between fermionic holons,
respectively, via hybridization fluctuations. Such two kinds of
pairing correlations result in multi-gap unconventional
superconductivity around the antiferromagnetic quantum critical
point of the slave-fermion theory, where $s-wave$ pairing is not
favored generically due to strong correlations. We show that the
ratio between each superconducting gap for conduction electrons
$\Delta_{c}$ and holons $\Delta_{f}$ and the transition
temperature $T_{c}$ is $2\Delta_{c} / T_{c} \sim 9$ and
$2\Delta_{f} / T_{c} \sim \mathcal{O}(10^{-1})$, remarkably
consistent with $CeCoIn_{5}$. A fingerprint of the Kondo mechanism
is emergence of two kinds of resonance modes in not only spin but
also charge fluctuations, where the charge resonance mode at an
antiferromagnetic wave vector originates from $d-wave$ pairing of
spinless holons. We discuss how the Kondo fluctuation theory
differs from the spin fluctuation approach.
\end{abstract}

\pacs{74.20.-z, 74.20.Mn, 71.10.Hf, 71.10.-w}

\maketitle

\section{Introduction}

Superconductivity from non-Fermi liquids has been one of the
central problems in modern condensed matter physics, associated
with high T$_{c}$ cuprates \cite{PALee_Review} and heavy-fermion
quantum critical points (QCPs) \cite{HF_Review}, where the theory
of superconductivity$-$BCS (Bardeen-Cooper-Schrieffer) mechanism
\cite{BCS} needs to be generalized, depending on the nature of the
normal state. When phonons are introduced in normal metals
described by Fermi liquid, where Coulomb interactions are screened
to become local allowing coherent electrons, there appear
attractive interactions between quasiparticles within the time
scale given by the Debye frequency associated with relaxation of
ions. Then, such attractions give rise to an instability of the
whole Fermi surface, causing superconductivity. On the other hand,
effective interactions between electrons are not screened
completely and still long-ranged around QCPs, causing incoherent
electron excitations and showing deviation of Fermi liquid
physics. A natural question is what happens if phonon excitations
are introduced into such a non-Fermi liquid metal. Will attractive
interactions emerge against such long-range interactions? Although
fully self-consistent diagrammatic analysis has not been performed
yet as far as we know, it would not be easy to generate such
attractions due to phonons.

The first point is on the nature of the non-Fermi liquid phase,
the hallmark of which is beyond the $T^{2}$ electrical
resistivity, for example, a typical $T$-linear behavior in various
heavy-fermion critical metals. A phenomenological description was
proposed, so called the marginal Fermi liquid ansatz, where
electron quasiparticles decay into bunch of particle-hole
excitations, regarded as an example of orthogonality catastrophe
\cite{MFL_Varma}. One possible origin is quantum criticality,
where scattering with critical fluctuations gives rise to the
$\omega \ln \omega$ self-energy with frequency $\omega$. Then, the
final question is whether such critical fluctuations as the source
of non-Fermi liquid physics will cause superconducting instability
in marginal Fermi liquid, where phonon excitations are not
expected to play an important role around QCPs.

The so called spin fluctuation scenario has been regarded as the
standard model for superconductivity out of non-Fermi liquids,
where the quantum critical normal state is described by the
Hertz-Moriya-Millis theory with the dynamical exponent $z$
\cite{HMM}, resulting in the temperature-linear resistivity in two
dimensions when $z = 2$ and effective interactions are oscillatory
in space, allowing unconventional pairing beyond s-wave
\cite{Chubukov_Review,Pines_Review}. This Fermi-liquid based
theory is quite parallel with the BCS theory, where phonons are
replaced with antiferromagnetic spin fluctuations and Migdal
theorem \cite{Chubukov_Vertex} holds in both mechanism. A
fingerprint of this non-phonon mechanism is emergence of spin
resonance modes at an antiferromagnetic wave vector in the
superconducting phase, analogous with line-width narrowing of the
phonon spectrum at frequency below twice of the superconducting
gap, actually measured in both high T$_{c}$ cuprates
\cite{Chubukov_Review} and heavy-fermion superconductors
\cite{Pines_Review}.

Recently, several heavy fermion compounds have been shown not to
follow the spin fluctuation scenario
\cite{GR,TR,dHvA,INS_Local_AF,Hall}. Anomalous thermodynamics such
as the divergent Gr\"{u}neisen ratio with an exponent $2/3$
\cite{GR} and non-Fermi liquid transport of temperature-linear
resistivity in three dimensions \cite{TR} turn out to be beyond
the description of the Fermi-liquid based theory
\cite{Kim_GR,Kim_TR}. Both divergence of the effective mass near
the QCP \cite{dHvA} and the presence of localized magnetic moments
at the transition towards magnetism \cite{INS_Local_AF} seem to
support a more exotic scenario. In addition, rather large entropy
and small magnetic moments in the antiferromagnetic phase may be
associated with antiferromagnetism out of a spin liquid Mott
insulator \cite{Senthil_Vojta_Sachdev}. Combined with the Fermi
surface reconstruction at the QCP \cite{dHvA,Hall}, this quantum
transition is assumed to show breakdown of the Kondo effect as an
orbital selective Mott transition \cite{Paul,Pepin,OSMT_HF}, where
only the f-electrons experience the metal-insulator transition.

The above discussion implies that superconductivity from such an
anomalous quantum critical metal is difficult to understand within
the spin fluctuation theory. Actually, superconductivity around
the antiferromagnetic QCP of $CeRhIn_{5}$ was claimed to be beyond
the spin fluctuation framework because this antiferromagnetic QCP
seems to be "local" associated with breakdown of the Kondo effect,
supported from the sub-linear-in-temperature electrical
resistivity and isotropic scattering emerging around the QCP, but
not in the heavy fermion phase \cite{TS_Nature}. Multi-gap
unconventional superconductivity was proposed in $CeCoIn_{5}$,
where large gap coexists with small gap associated with various
Fermi surfaces \cite{Multi_SC}, requiring a new kind of
theoretical framework for superconductivity around the QCP.

In the theoretical point of view two kinds of heavy-fermion QCPs
were proposed, where nature of spin dynamics is at the heart of
heavy fermion quantum criticality \cite{Coleman_Pepin_Si}. The
RKKY (Ruderman-Kittel-Kasuya-Yosida) induced antiferromagnetic QCP
is nothing but the Stoner instability of heavy quasiparticles, and
only small pieces of Fermi surface become critical via nesting.
Superconductivity out of this QCP is described by the spin
fluctuation mechanism. On the other hand, the Kondo induced QCP
leads the whole Fermi surface to be critical, associated with
formation of heavy quasiparticles. Recently, a dynamical
mean-field theory study has shown that the Kondo induced QCP may
be identified with an orbital selective Mott transition
\cite{OSMT_HF}, where spins become localized in the
antiferromagnetic phase in contrast with the RKKY mechanism.
Importance of Gutzwiller projection was emphasized in the Mott
limit of one-band Hubbard model \cite{PWAnderson}. In this respect
the weak coupling approach is difficult to apply to the strong
coupling problem. An important question is to develop the field
theoretic manipulation for Gutzwiller projection.

In this paper we propose Kondo fluctuation mechanism of
superconductivity, differentiated from the spin fluctuation
theory. Based on the U(1) slave-fermion representation of an
effective Anderson lattice model, where localized spins are
described by the Schwinger boson theory \cite{Sachdev_SpN} and
hybridization or Kondo fluctuations weaken antiferromagnetic
correlations of localized spins, we found an antiferromagnetic QCP
from an antiferromagnetic metal to a heavy fermion metal in our
recent study \cite{Kim_Jia}. The Kondo induced antiferromagnetic
QCP was shown to be described by both conduction electrons and
fermionic holons interacting with critical spin fluctuations given
by deconfined bosonic spinons with a spin quantum number $1/2$.
Surprisingly, such critical modes turned out to be described by
the dynamical exponent $z = 3$, giving rise to the well known
non-Fermi liquid physics such as the divergent Gr\"{u}neisen ratio
with an exponent $2/3$ \cite{Kim_GR} and temperature-linear
resistivity in three dimensions \cite{Kim_TR}. We find that the $z
= 3$ antiferromagnetic QCP becomes unstable against
superconductivity, where critical spinon excitations give rise to
pairing correlations between conduction electrons and between
fermionic holons, respectively, via hybridization fluctuations.
Such two kinds of pairing correlations result in multi-gap
unconventional superconductivity around the antiferromagnetic QCP
of the slave-fermion theory, where $s-wave$ pairing is not favored
generically due to strong correlations. We show that the ratio
between each superconducting gap for conduction electrons
$\Delta_{c}$ and holons $\Delta_{f}$ and the transition
temperature $T_{c}$ is $2\Delta_{c} / T_{c} \sim 9$ and
$2\Delta_{f} / T_{c} \sim \mathcal{O}(10^{-1})$, remarkably
consistent with $CeCoIn_{5}$ \cite{Multi_SC}. A fingerprint of the
Kondo mechanism is emergence of two kinds of resonance modes in
not only spin but also charge fluctuations, where the charge
resonance mode at an antiferromagnetic wave vector originates from
$d-wave$ pairing of spinless holons. We argue uniqueness and
robustness of the Kondo fluctuation mechanism, comparing with
other scenarios based on hybridization fluctuations such as the
valance-fluctuation \cite{Miyake}, resonating-valance-bond (RVB)
\cite{Senthil_Vojta_Sachdev}, and two channel SU(2) slave-boson
\cite{Coleman} theories.

\section{U(1) slave-fermion theory of Anderson lattice model}

\subsection{U(1) slave-fermion representation of an effective Anderson lattice model}

We start from an effective Anderson lattice model \bqa && H_{ALM}
= H_{c} + H_{f} + H_{Kondo} + H_{RKKY} , \nn && H_{c} = -
t\sum_{\langle ij\rangle}(c_{in\sigma}^{\dagger}c_{jn\sigma} +
H.c.) - \mu \sum_{i}c_{in\sigma}^{\dagger}c_{in\sigma} , \nn &&
H_{f} = - \alpha t\sum_{\langle
ij\rangle}(d_{in\sigma}^{\dagger}d_{jn\sigma} + H.c.) +
\epsilon_{f}\sum_{i}d_{in\sigma}^{\dagger}d_{in\sigma} , \nn &&
H_{Kondo} = V\sum_{i}(c_{in\sigma}^{\dagger}d_{in\sigma} + H.c.) ,
\nn && H_{RKKY} = \frac{J}{N}\sum_{\langle i j
\rangle}\vec{S}_{i}\cdot\vec{S}_{j} , \eqa exhibiting competition
between hybridization fluctuations $H_{Kondo}$ and
antiferromagnetic correlations of localized spins $H_{RKKY}$,
where the large-$U$ limit for localized orbitals is taken into
account. In particular, RKKY interactions are modelled as
effective exchange interactions between localized spins. We also
assume the presence of week hopping integrals for localized
electrons, denoted by $\alpha \ll 1$. Although the hybridization
term gives rise to both RKKY interactions in its fourth order
($\sim V^{4}/U^{3}$) and hopping integrals in its second order
($\sim V^{2}/U$) \cite{Hewson_Book}, we regard this effective
Anderson model as an emergent model of the intermediate energy
scale in the renormalization group sense. Here, $\sigma =
\uparrow, \downarrow$ represents SU(2) spin and $n = 1, ..., N$
expresses the number of flavors, allowing us to analyze this model
in a systematic way.

Expressing an electron field in a localized orbital as \bqa &&
d_{in\sigma} = f_{i}^{\dagger}b_{in\sigma} , \eqa where $f_{i}$
carries only charge, called holon, and $b_{in\sigma}$ does only
spin, called spinon, the large-$U$ limit in the localized orbital
is expressed as the so called single occupancy constraint, \bqa &&
\sum_{n = 1}^{N} \sum_{\sigma = \uparrow, \downarrow}
b_{in\sigma}^{\dagger}b_{in\sigma} + f_{i}^{\dagger}f_{i} = 2 S N
. \eqa In the final stage of calculation we will consider $S =
1/2$ and $N = 1$.

Inserting the U(1) slave-fermion representation into the RKKY
term, we take \bqa && \frac{J}{N}\sum_{\langle
ij\rangle}\vec{S}_{i}\cdot\vec{S}_{j} = - \frac{J}{N}
\sum_{\langle ij\rangle} (\epsilon_{\alpha\beta}
b_{in\alpha}^{\dagger}b_{jn\beta}^{\dagger} )
(\epsilon_{\gamma\delta} b_{im\gamma}b_{jm\delta} ) \nn &&
\rightarrow \sum_{\langle ij\rangle} \Bigl\{
\frac{N}{J}|\Delta_{ij}|^{2} - (
\Delta_{ij}^{*}\epsilon_{\sigma\sigma'}b_{in\sigma}b_{jn\sigma'} +
H.c. ) \Bigr\} \eqa for antiferromagnetic correlations, where
$\Delta_{ij}$ capture spin-singlet excitations. In the same way we
see \bqa && - \alpha t\sum_{\langle
ij\rangle}(d_{in\sigma}^{\dagger}d_{jn\sigma} + H.c.) \nn && = -
\alpha t\sum_{\langle ij\rangle}
(b_{in\sigma}^{\dagger}f_{i}f_{j}^{\dagger}b_{jn\sigma} + H.c.)
\nn && \rightarrow \alpha t\sum_{\langle ij\rangle} \Bigl\{
(\chi_{ij}^{b*}\chi_{ij}^{f} + H.c.) -
(f_{i}\chi_{ij}^{b*}f_{j}^{\dagger} + H.c.) \nn && -
(b_{in\sigma}^{\dagger}\chi_{ij}^{f}{b}_{jn\sigma} + H.c.) \Bigr\}
, \eqa where $\chi_{ij}^{b}$ keep hopping fluctuations for holons
and $\chi_{ij}^{f}$ take ferromagnetic correlations.

Based on Eqs. (4) and (5), we obtain an effective Lagrangian in
the U(1) slave-fermion representation of the Anderson lattice
model \bqa && Z =
\int{Dc_{in\sigma}Db_{in\sigma}Df_{i}D\Delta_{ij}D\chi_{ij}^{b}D\chi_{ij}^{f}D\lambda_{i}}
e^{- \int_{0}^{\beta}{d\tau} L} , \nn && L = L_{c} + L_{f} + L_{b}
+ L_{V} + L_{0} , \nn && L_{c} =
\sum_{i}c_{in\sigma}^{\dagger}(\partial_{\tau} - \mu)c_{in\sigma}
- t\sum_{\langle ij\rangle}(c_{in\sigma}^{\dagger}c_{jn\sigma} +
H.c.) , \nn && L_{f} = \sum_{i} f_{i}^{\dagger}(\partial_{\tau} +
i\lambda_{i})f_{i} + \alpha t\sum_{\langle ij\rangle}
(f_{j}^{\dagger}\chi_{ij}^{b*}f_{i} + H.c.) , \nn && L_{b} =
\sum_{i} b_{in\sigma}^{\dagger}(\partial_{\tau} + \epsilon_{f} +
i\lambda_{i})b_{in\sigma} - \alpha t\sum_{\langle ij\rangle} (
b_{in\sigma}^{\dagger}\chi_{ij}^{f}b_{jn\sigma} \nn && + H.c.) - J
\sum_{\langle ij\rangle} (
\Delta_{ij}^{*}\epsilon_{\sigma\sigma'}b_{in\sigma}b_{jn\sigma'} +
H.c. ) , \nn && L_{V} = V \sum_{i}
(c_{in\sigma}^{\dagger}b_{in\sigma}f_{i}^{\dagger} + H.c.) , \nn
&& L_{0} = \alpha t\sum_{\langle ij\rangle}
(\chi_{ij}^{b*}\chi_{ij}^{f} + H.c.) + N J \sum_{\langle
ij\rangle} |\Delta_{ij}|^{2} \nn && - i \sum_{i} 2 N S\lambda_{i}
, \eqa where the hybridization term $V$ competes with the
antiferromagnetic correlation term $J$ for localized electrons,
modelled as the nearest neighbor spin-exchange interaction.
$L_{c}$ describes dynamics of conduction electrons $c_{in\sigma}$,
where $\mu$ and $t$ are their chemical potential and kinetic
energy, respectively. $L_{f}$ and $L_{b}$ govern dynamics of
localized electrons, decomposed with fermionic holons $f_{i}$ and
bosonic spinons $b_{in\sigma}$, where local antiferromagnetic
correlations $\Delta_{ij}$ are introduced in the Sp(N)
representation for the spin-exchange term $J$ with an index $n =
1, ..., N$ \cite{Sachdev_SpN} and an almost flat band with $\alpha
\ll 1$ is allowed \cite{Pepin} to describe hopping of holons
$\chi_{ij}^{b}$ and spinons $\chi_{ij}^{f}$, respectively.
$\epsilon_{f}$ is an energy level for the flat band, and
$\lambda_{i}$ is a Lagrange multiplier field to impose the
slave-fermion constraint. $L_{V}$ is the hybridization term,
involving conduction electrons, holons, and spinons. $L_{0}$
represents condensation energy with $N = 1$ and $S = 1/2$ in the
physical case.

In the decoupling limit of $V \rightarrow 0$ the slave-fermion
Lagrangian is reduced to two decoupled sectors for conduction
electrons and spinons, where ferromagnetic correlations
$\chi_{ij}^{f}$ vanish due to $\langle f_{i}^{\dagger}f_{i}
\rangle = 0$ in the spinon sector, recovering the Schwinger-boson
theory for the half filled quantum antiferromagnet
\cite{Sachdev_SpN} \bqa && Z = \int{Dc_{in\sigma}Db_{in\sigma}
D\Delta_{ij} D\lambda_{i}} e^{- \int_{0}^{\beta}{d\tau} (L_{c} +
L_{b})} , \nn && L_{c} =
\sum_{i}c_{in\sigma}^{\dagger}(\partial_{\tau} - \mu)c_{in\sigma}
- t\sum_{\langle ij\rangle}(c_{in\sigma}^{\dagger}c_{jn\sigma} +
H.c.) , \nn && L_{b} = \sum_{i}
b_{in\sigma}^{\dagger}(\partial_{\tau} + \epsilon_{f} +
i\lambda_{i})b_{in\sigma} \nn && - \sum_{\langle ij\rangle} (
\Delta_{ij}^{*}\epsilon_{\sigma\sigma'}b_{in\sigma}b_{jn\sigma'} +
H.c. ) \nn && + \frac{N}{J} \sum_{\langle ij\rangle}
|\Delta_{ij}|^{2} - i \sum_{i} 2 N S\lambda_{i} . \eqa Actually,
this is our starting point for the description of localized spins
instead of itinerant electrons in the Hertz-Moriya-Millis theory.
In this respect the present problem generalizes the
Schwinger-boson theory, turning on hybridization fluctuations to
cause "hole doping" in the localized band, represented by
fermionic holons $f_{i}$. Particulary, hybridization fluctuations
give rise to ferromagnetic correlations via effective hole doping,
weakening antiferromagnetic correlations $\Delta_{ij}$ and
destroying the antiferromagnetic order $\langle b_{in\sigma}
\rangle = 0$.



\subsection{$z = 3$ antiferromagnetic quantum critical metal}

In the recent study we performed the mean-field analysis with
uniform hopping $\chi_{ij}^{f(b)} \rightarrow \chi_{f(b)}$,
pairing $\Delta_{ij} \rightarrow \Delta$, and chemical potential
$i\lambda_{i} \rightarrow \lambda$, and found the slave-fermion
mean-field phase diagram for the Anderson lattice model (Fig. 1)
\cite{Kim_Jia}. The antiferromagnetic long range order turns out
to vanish at the critical hybridization strength $V_{c}$, but
short range antiferromagnetic correlations still exist at the QCP.
In the antiferromagnetic phase ($V \ll V_{c}$) band hybridization
is allowed, but the area of the Fermi surface is small,
proportional to $\delta$, the density of conduction electrons,
because the effective chemical potential of holons is almost on
the top of the holon band and the density of holons is vanishingly
small. Enhancing the hybridization coupling constant ($V >
V_{c}$), the holon chemical potential shifts to the lower part,
filling holons into the flat band and causing heavy fermions. In
this description the heavy fermion transition at finite
temperatures turns into crossover, where the crossover temperature
$T_{FL}$ is given by gap of spinon excitations $T_{FL} \sim
\xi_{s}^{-1}$ with the correlation length $\xi_{s} = [(\lambda - 2
d \alpha t \chi_{f})^{2} - (2 d \Delta )^{2}]^{-1/2}$ since
scattering of conduction electrons and holons with spinon
fluctuations is suppressed below this temperature allowing Fermi
liquid physics.

Fluctuation-corrections are taken into account for quantum
critical physics in the Eliashberg framework, where vertex
corrections are neglected \cite{Paul,Pepin}. Our main discovery
was that dynamics of spinon fluctuations is described by $z = 3$
critical theory due to Landau damping of electron-holon
polarization above an intrinsic energy scale $E^{*}$, while by $z
= 1$ O(4) nonlinear $\sigma$ model below $E^{*}$ \cite{Kim_Jia}.
The energy scale $E^{*} \propto \alpha D (q^{*}/k_{F}^{c})^{3}$
originates from the mismatch $q^{*} = |k_{F}^{f} - k_{F}^{c}|$ of
the Fermi surfaces of the conduction electrons $k_{F}^{c}$ and
holons $k_{F}^{f}$, shown to vary from ${\cal O}(10^{0})$ $mK$ to
${\cal O}(10^{2})$ $mK$ \cite{Paul,Pepin}. Actually, inserting the
Landau damping self-energy $\Pi_{b}(q,i\Omega) = \gamma_{b}
\frac{|\Omega|}{q}$ with the damping coefficient $\gamma_{b} =
\frac{\pi}{2} \frac{V^{2}\rho_{c}}{v_{F}^{f}}$ into the spinon's
full propagator, where $\rho_{c}$ is the density of states for
conduction electrons and $v_{F}^{f}$ is the holon velocity, we
find their $z = 3$ dynamics \bqa && \Im G_{b}(q,\Omega) \approx -
\frac{\gamma}{2\gamma_{b}} \frac{ \gamma \Omega q }{ q^{6} +
\gamma^{2} \Omega^{2} } \eqa with $\gamma \equiv (2\gamma_{b})(2 d
\Delta/v_{s}^{2})$, where $v_{s} = \sqrt{2 [\alpha t \chi_{f}
(\lambda - 2 d \alpha t \chi_{f}) + (2 d \Delta^{2} )]}$ is the
velocity of spinons. Then, the correlation-length exponent is
given by the usual mean-field value $\nu = 1/2$ since the critical
theory is above its upper critical dimension in $d = 3$
\cite{Dirre}.

Both anomalous thermodynamics and non-Fermi liquid transport
result from the $z = 3$ quantum criticality. The so called
Gr\"{u}neisen ratio, the ratio between the thermal expansion
parameter and specific heat coefficient, diverges with an exponent
$\frac{1}{\nu z} = \frac{2}{3}$, where $\nu$ is the correlation
length exponent \cite{GR,Kim_GR}. The electrical resistivity
displays the $T$-linear behavior in three spatial dimensions
\cite{TR,Kim_TR}, different from the $z = 2$ spin-density-wave
theory ($\sim T^{3/2}$). An important result of the $z = 3$
antiferromagnetic QCP in the slave-fermion theory is that the
uniform dynamic spin susceptibility diverges with an exponent
$2/3$, similar to an experiment \cite{INS_Local_AF}. Of course,
the staggered spin susceptibility diverges as it should be due to
the antiferromagnetic instability. Divergence of the uniform spin
susceptibility is an inevitable response from the $z = 3$
antiferromagnetic QCP. As a result, the $z = 3$ antiferromagnetic
QCP should be distinguished from the $z = 2$ spin-density-wave
theory, where physical response functions for the $z = 3$
antiferromagnetic QCP are summarized in Table I.


\begin{table}[ht]
\begin{tabular}{ccccc}
\; & $z$ \& $\nu$ \; & $\Gamma (T)$ \; & $\chi (T)$ \; & $\rho
(T)$ \; \nn \hline SF QCP \; & 3 \& 1/2 \; \; & $T^{-2/3}$ \; \; &
$T^{-2/3}$ \; \; & $T \ln (2T/E^{*})$ \; \;  \nn \hline
\end{tabular}
\caption{Scaling of Gr\"{u}neisen ratio $\Gamma (T)$, uniform spin
susceptibility $\chi (T)$, and resistivity $\rho (T)$ with
dynamical $z$ and correlation-length $\nu$ exponents in $d=3$ for
the slave-fermion theory}
\end{table}

\begin{figure}[t]
\vspace{6cm} \includegraphics{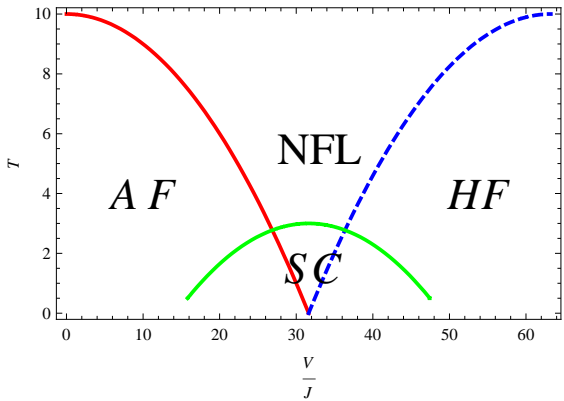} \caption{ (Color online) Schematic
phase diagram for hybridization-fluctuation-induced $d-wave$
superconductivity around the $z = 3$ AF QCP with an AF transition
temperature (red thick), crossover temperature to the heavy
fermion (HF) phase (blue dashed), superconducting (SC) transition
temperature (green thick), and non-Fermi liquid (NFL), where both
the red thick and blue dashed lines were found in Ref.
\cite{Kim_Jia}.} \label{fig1} \vspace{-0.5cm}
\end{figure}

\subsection{Discussion on self-consistency in the slave-fermion theory}

One cautious physicist may suspect existence of the $z = 3$
antiferromagnetic QCP because such a QCP is based on deconfinement
of fractionalized excitations called spinons and holons. If such
elementary excitations should be confined from some
non-perturbative effects, for example, due to magnetic monopole
excitations, the present description becomes illusive. In this
subsection we discuss that the effective U(1) gauge theory in the
slave-fermion representation allows deconfinement of such
fractionalized excitations at its QCP, thus this kind of theory is
self-consistent in itself.

Before discussing deconfinement of slave-particles, we would like
to explain the motivation of the slave-particle representation.
For clarity, suppose the one-band Hubbard model at half filling
without perfect nesting. It is believed that the spin fluctuation
approach is applied to the $u/t < 1$ regime while the Gutswiller
projection should be introduced in $u/t > 1$, simulated with the
slave-particle representation. In the latter case the interaction
coupling constant $u$ was argued to increase more and more, going
to an infinite coupling fixed point \cite{PWAnderson}. Of course,
since this is beyond the perturbative regime and the usual
one-loop renormalization group analysis does not work, this
statement is just one claim based on the method of canonical
transformation. However, it seems to be true that the Fermi-liquid
based approach is difficult to simulate the Gutzwiller projection.
Actually, nobody did not reach the Mott transition regime based on
the Fermi-liquid based approach, where the whole Fermi surface
becomes critical while the Fermi-liquid based theory exhibits
instability of only some parts of the Fermi surface. In the
present context the Kondo effect is not reached yet based on the
spin fluctuation approach as far as we know. This is the strong
motivation for the slave-particle representation.

To check whether the slave-particle theory is self-consistent or
not means to understand whether deconfinement is allowed or not
beyond the perturbative analysis. Here, "beyond the perturbative
analysis" expresses that magnetic monopole excitations are
introduced, allowed in the lattice U(1) gauge theory
\cite{Confinement_Intro}. Their condensation gives rise to
confinement of slave-particles, then the present approach loses
its physical implication.

It has been known that the pure lattice U(1) gauge theory without
matter fields allows deconfinement in three spatial dimensions
\cite{Confinement_Intro}. More precisely, there is the
confinement-deconfinement transition varying the coupling
constant, here the internal "electric" charge. In the
deconfinement phase introduction of matters strengthens
deconfinement, and slave-particles appear as elementary
excitations. An important question is what happens in the
confinement phase of the pure gauge theory if we introduce matter
fields. This has been regarded as an important issue in the gauge
theory approach to strongly correlated electrons. Recently, some
reliable arguments have been made.

When matter fields are gapped, confinement survives, of course.
The question is what happens when matter fluctuations are critical
or gapless. Hermele et al. claimed that magnetic monopole
excitations can be suppressed when there are plenty of flavors for
Dirac fermions in QED$_{3}$ (quantum electrodynamics in two space
and one time dimensions) \cite{Deconfinement_ASL}. More precisely,
they showed that the scaling dimension of the monopole excitation
operator is given by the flavor number $N$ of Dirac fermions at
the infrared stable fixed point so called the algebraic spin
liquid. At the fixed point of QED$_{3}$ they calculated energy of
one magnetic monopole, proportional to the fermion flavor number
$N$. Based on the state-operator correspondence of the conformal
field theory, such an energy is identified with the scaling
dimension of the magnetic monopole insertion operator. Since it is
proportional to $N$, monopole excitations become irrelevant in the
large $N$ limit. This relativistic study was extended to the
non-relativistic case, where there is a Fermi surface of spinons
\cite{Deconfinement_U1SL_SS,Deconfinement_U1SL_KS}. Since there
are plenty of fermions around the Fermi surface, one may expect
that deconfinement always occurs. Actually, it was argued that
deconfinement indeed happens at the spin liquid fixed point
\cite{Deconfinement_U1SL_SS}. A similar result was also obtained
in the case of bosonic matters \cite{Deconfinement_Boson}.

In the present U(1) slave-fermion gauge theory we have two kinds
of critical matters involved with the internal U(1) gauge charge.
These are holons with a Fermi surface and gapless spinons at the
antiferromagnetic QCP. In this respect deconfinement is allowed,
thus the present theoretical framework is self-consistent at least
around the QCP.

Finally, we would like to mention one of the main successes in
this approach. Applying the slave-particle representation to the
multi-channel Kondo problem, one finds the scaling solution of a
power law within the so called non-crossing approximation,
identified with a non-Fermi liquid fixed point due to
over-screening. Actually, this physics turns out to coincide with
the exact method, the conformal field theory \cite{NCA_CFT}.

\section{Superconductivity from a non-Fermi liquid metal}

\subsection{Superconducting instability of the $z = 3$
antiferromagnetic quantum critical point}

First, we show that the $z = 3$ antiferromagnetic QCP becomes
unstable against unconventional superconductivity, evaluating
particle-particle scattering vertices for both conduction
electrons and holons with subscripts $c$ and $f$, respectively,
\bqa && \Phi_{cc}(i\Omega) = - V^{2} \frac{1}{\beta} \sum_{i\nu}
\sum_{l} \Phi_{ff}(i\Omega+i\nu) F_{b}(l,i\nu) \nn &&
G_{f}(k_{F}^{c} +l, i\Omega+i\nu) G_{f}(-k_{F}^{c} -l,
-i\Omega-i\nu) , \nn && \Phi_{ff}(i\Omega) = - 2N V^{2}
\frac{1}{\beta} \sum_{i\nu} \sum_{l} \Phi_{cc}(i\Omega+i\nu)
F_{b}(l,i\nu) \nn && G_{c}(k_{F}^{f} +l, i\Omega+i\nu)
G_{c}(-k_{F}^{f} -l, -i\Omega-i\nu) . \eqa
$\Phi_{cc(ff)}(i\Omega)$ is the particle-particle t-matrix for
conduction electrons (holons) and \bqa && G_{c(f)}(k,i\omega) =
\frac{1}{i\omega - (\epsilon_{k}^{c(f)} - \mu_{c(f)}) -
\Sigma_{n}^{c(f)}(i\omega)} \nonumber \eqa is the normal Green's
function for conduction electrons (holons), where
$\epsilon_{k}^{c} = - 2 t (\cos k_{x} + \cos k_{y} + \cos k_{z})$
and $\epsilon_{k}^{f} = - \alpha \chi_{b} \epsilon_{k}^{c}$ are
fermion dispersions, and $\mu_{c} = \mu$ and $\mu_{f} = - \lambda$
are their chemical potentials. $\Sigma_{n}^{c(f)}(i\omega)$ is the
normal self-energy of conduction electrons (holons),
self-consistently found in the Eliashberg approximation
\cite{Kim_Jia}. \bqa && F_{b}(q,i\Omega) \nn && = \frac{-
\frac{\epsilon_{q}^{b}}{t} \Delta}{-(i\Omega)^{2} +
[\epsilon_{q}^{b} + \epsilon_{f} + \lambda +
\Pi_{b}(q,i\Omega)]^{2} - (\epsilon_{q}^{b}/t)^{2}\Delta^{2}}
\nonumber \eqa is an anomalous propagator for spinons due to their
pairing correlations, shown in $H_{RKKY}$ of Eq. (4), where
$\epsilon_{q}^{b} = \alpha \chi_{f} \epsilon_{q}^{c}$ is the
spinon bare dispersion. $\Pi_{b}(q,i\Omega)$ is the normal
self-energy given by the Landau damping form, as discussed before.
The presence of the anomalous spinon propagator or
antiferromagnetic correlations is an important ingredient for the
Kondo fluctuation mechanism, discussed in more detail later (Fig.
2). The negative sign in the right hand side implies that $s-wave$
superconductivity is prohibited as expected due to strong
correlations.


\begin{figure}[t]
\vspace{0cm}
\includegraphics[width=0.6\textwidth]{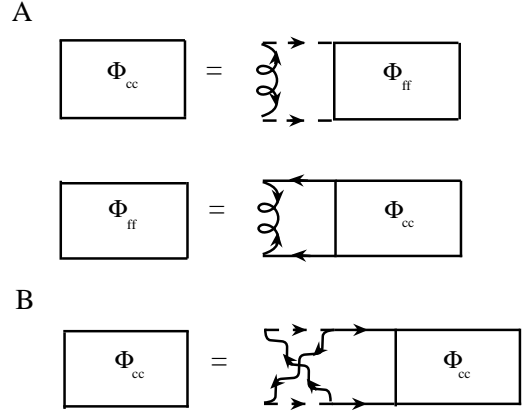} \vspace{-2cm} \caption{ A. Coupled
particle-particle t-matrices for both conduction electrons and
holons in the slave-fermion theory, where the thick line is the
electron's Green function, the dashed line is the holon's Green
function, and the coiling line is the anomalous spinon's Green
function. B. A particle-particle t-matrix for conduction electrons
in the slave-boson theory, where the thick line is the electron's
Green function, the dashed line is the spinon's Green function,
and the wavy line is the normal holon's Green function.}
\label{fig2} \vspace{-0.5cm}
\end{figure}

In the long wave length and low energy limits the anomalous spinon
Green's function can be written as \bqa && \Im F_{b}(q,\Omega) =
\frac{\gamma}{2\gamma_{b}} \frac{\gamma \Omega q}{q^{6} +
\gamma^{2} \Omega^{2}} . \nonumber \eqa Inserting this expression
into the above and performing momentum integration with the ansatz
of $d-wave$ pairing, we obtain \bqa && \Phi_{cc}(i\Omega) \approx
\frac{ \mathcal{C}_{c}^{2} }{2} \frac{1}{\beta} \sum_{i\nu} \ln
\Bigl( \frac{\Omega_{c} + |\nu-\Omega|}{ |\nu-\Omega|} \Bigr)
\frac{\Phi_{ff}(i\nu)}{ |\nu + i \Sigma_{n}^{f}(i\nu)| } , \nn &&
\Phi_{ff}(i\Omega) \approx 2N \frac{ \mathcal{C}_{f}^{2} }{2}
\frac{1}{\beta} \sum_{i\nu} \ln \Bigl( \frac{\Omega_{c} +
|\nu-\Omega|}{ |\nu-\Omega|} \Bigr) \frac{\Phi_{cc}(i\nu)}{ |\nu +
i \Sigma_{n}^{c}(i\nu)| } , \nn \eqa where $z > 1$ ($z = 3$, here)
quantum criticality allows the local form for spinon fluctuations
with their cutoff frequency $\Omega_{c}$, and the coupling
constants are given by $\mathcal{C}_{c(f)}^{2} = \frac{ 4 d \pi
V^{2} \Delta}{ (2\pi)^{3} z v_{s}^{2} v_{F}^{f(c)} }$ with the
spinon velocity
$v_{s} = \sqrt{2 [\alpha t \chi_{f} (\lambda - 2 d \alpha t
\chi_{f}) + (2 d \Delta^{2} )]}$ \cite{Kim_Jia} and holon
(conduction electron) Fermi velocity $v_{F}^{f(c)}$.

Absence of quasiparticles at the $z = 3$ antiferromagnetic QCP is
seen from the following fermion self-energies \bqa && i
\Sigma_{n}^{c}(i\omega) = g_{c}^{2} \omega \ln
\frac{\Omega_{c}}{|\omega|} , ~~~~~ i \Sigma_{n}^{f}(i\omega) =
g_{f}^{2} \omega \ln \frac{\Omega_{c}}{|\omega|} ,   \eqa where
$g_{c}^{2} = \frac{ d V^{2} \Delta}{6 \pi^{2} v_{s}^{2}
v_{F}^{f}}$ and $g_{f}^{2} = 2N \frac{ d V^{2} \Delta}{6\pi^{2}
v_{s}^{2} v_{F}^{c}}$ \cite{Kim_GR,Kim_TR}. Although this
corresponds to the marginal Fermi liquid ansatz, its mechanism in
the strong coupling approach differs from the spin fluctuation
scenario. Inserting these non-Fermi liquid self-energies, Eq. (10)
can be written as follows
\bqa && \Phi_{cc}(i\Omega) \approx \mathcal{C}_{c}^{2}
\int_{T_{c}}^{\infty} d \nu \frac{\Phi_{ff}(i\nu)}{ \nu \Bigl(1 +
g_{f}^{2} \ln \frac{\Omega_{c}}{ \nu } \Bigr) } \ln
\frac{\Omega_{c}}{\sqrt{|\nu^{2} - \Omega^{2}|}} , \nn &&
\Phi_{ff}(i\Omega) \approx 2N \mathcal{C}_{f}^{2}
\int_{T_{c}}^{\infty} d \nu \frac{\Phi_{cc}(i\nu)}{ \nu \Bigl(1 +
g_{c}^{2} \ln \frac{\Omega_{c}}{ \nu } \Bigr) } \ln
\frac{\Omega_{c}}{\sqrt{|\nu^{2} - \Omega^{2}|}} , \nn  \eqa where
finite temperature effects are introduced as the lower cutoff
approximately \cite{Chubukov_Eliashberg_Theory}. Following the
procedure of Ref. \cite{Chubukov_Eliashberg_Theory}, we find \bqa
&& T_{c} \approx \Omega_{c} e^{- \frac{\pi}{\sqrt{\sqrt{2N }
\mathcal{C}_{c} \mathcal{C}_{f} } } } \eqa in the "BCS" limit
$g_{c(f)}^{2} \ll 1$. An important lesson in this expression is
that the $1/\sqrt{\mathcal{C}_{c} \mathcal{C}_{f}} \propto 1/V$
factor in the exponential appears instead of $1/V^{2}$, associated
with the absence of quasiparticles. Using appropriate parameters
shown to fit thermodynamics of $YbRh_{2}Si_{2}$ qualitatively well
\cite{Kim_GR}, we see that $T_{c}$ varies from
$\mathcal{O}(10^{0}) K$ to $\mathcal{O}(10^{1}) K$ depending on
$10 K \leq \Omega_{c} \leq 30 K$, consistent with
$Ce(Co,Rh)In_{5}$ \cite{TS_Nature}.

\subsection{Kondo fluctuation mechanism : Multi-gap
superconductivity}


To understand the $d-wave$ superconductivity around the $z = 3$
deconfined antiferromagnetic QCP, we develop an Eliashberg theory
\cite{Chubukov_Review} for the hybridization-induced
superconductivity. The Luttinger-Ward functional can be
constructed as $Y_{LW} = Y_{LW}^{N} + Y_{LW}^{S}$ with \bqa &&
Y_{LW}^{N} = 2 N V^{2} \frac{1}{\beta} \sum_{i\Omega } \sum_{q}
\frac{1}{\beta} \sum_{i\omega } \sum_{k} G_{c}(k+q,i\omega
+i\Omega ) \nn && G_{b}(q,i\Omega ) G_{f}(k,i\omega ) , \nn &&
Y_{LW}^{S} = - 2 N V^{2} \frac{1}{\beta} \sum_{i\Omega} \sum_{q}
\frac{1}{\beta} \sum_{i\omega } \sum_{k} F_{c}(k+q,i\omega
+i\Omega ) \nn && F_{b}(q,i\Omega ) F_{f}(k,i\omega ) , \eqa where
$Y_{LW}^{N}$ is for normal self-energies with each normal Green's
function and $Y_{LW}^{S}$ is for anomalous self-energies with each
anomalous propagator \cite{Dirre}. \bqa && G_{c(f)}(k,i\omega) \nn
&& = \frac{i\omega - \Sigma_{n}^{c(f)}(i\omega) +
(\epsilon_{k}^{c(f)} - \mu_{c(f)})}{[i\omega -
\Sigma_{n}^{c(f)}(i\omega)]^{2} - (\epsilon_{k}^{c(f)} -
\mu_{c(f)})^{2} - \Sigma_{p}^{c(f)2}(k,i\omega)} \nonumber \eqa is
the normal electron (holon) Green's function with the $d-wave$
pairing anomalous self-energy $\Sigma_{p}^{c(f)}(k,i\omega)$, and
$G_{b}(q,i\Omega)$ is the normal spinon propagator, where its
anomalous self-energy can be neglected as long as it is smaller
than its pairing order $\Delta$. \bqa && F_{c(f)}(k,i\omega) \nn
&& = \frac{\Sigma_{p}^{c(f)}(k,i\omega)}{[i\omega -
\Sigma_{n}^{c(f)}(i\omega)]^{2} - (\epsilon_{k}^{c(f)} -
\mu_{c(f)})^{2} - \Sigma_{p}^{c(f)2}(k,i\omega)} \nonumber \eqa is
the anomalous electron (holon) Green's function, and the anomalous
spinon propagator is the same as before because the presence of
antiferromagnetic correlations $\Delta$ allows us to neglect its
anomalous self-energy.

The electron and holon pairing self-energies are given by
\bqa \Sigma_{p}^{c}(k_{F}^{c},i\omega ) = \frac{ V^{2} }{2\pi
v_{F}^{f}} \frac{1}{\beta} \sum_{i\Omega }
\frac{\Sigma_{p}^{f}( i\Omega
)F_{b}(i\Omega-i\omega)}{\sqrt{\Bigl( \Omega + i \Sigma_{n}^{f}(
i\Omega ) \Bigr)^{2} + \Sigma_{p}^{f2}( i\Omega )}} , \nn
\Sigma_{p}^{f}(k_{F}^{f},i\omega) = \frac{ 2N V^{2} }{ 2\pi
v_{F}^{c}} \frac{1}{\beta} \sum_{i\Omega }
\frac{\Sigma_{p}^{c}( i\Omega
)F_{b}(i\Omega-i\omega)}{\sqrt{\Bigl( \Omega + i \Sigma_{n}^{c}(
i\Omega ) \Bigr)^{2} + \Sigma_{p}^{c2}( i\Omega )}} , \nn \eqa
where $d-wave$ pairing is assumed in the sign of the anomalous
fermion self-energy, and $F_{b}(i\Omega) = \int \frac{d^{d-1}
q_{\perp}}{(2\pi)^{d-1}} F_{b}(q_{\perp},i\Omega)$.
This expression is consistent with Eq. (10), justifying our
derivation of Eliashberg equations for pairing self-energies.

It is valuable to find the BCS limit of these equations
appropriate for the "weak" coupling case. We obtain coupled BCS
equations for electron and holon pairing order parameters \bqa &&
\Delta_{c} = \mathcal{B}_{c} \int_{0}^{\Omega_{c}} d \xi
\frac{\Delta_{f}}{\sqrt{\xi^{2} + \Delta_{f}^{2}}}
\tanh\frac{\sqrt{\xi^{2} + \Delta_{f}^{2}}}{2T} , \nn &&
\Delta_{f} = 2N \mathcal{B}_{f} \int_{0}^{\Omega_{c}} d \xi
\frac{\Delta_{c}}{\sqrt{\xi^{2} + \Delta_{c}^{2}}}
\tanh\frac{\sqrt{\xi^{2} + \Delta_{c}^{2}}}{2T} , \eqa where
$\mathcal{B}_{c(f)} = \mathcal{C}_{c(f)}^{2} \ln \Bigl(1 +
\frac{v_{s}^{2}\Omega_{c}^{2/3}}{m_{s}^{2}} \Bigr)$ with mass of
spinons $m_{s}^{2} \propto \sqrt{(\lambda - 2 d \alpha t
\chi_{f})^{2} - (2 d \Delta)^{2}}$ \cite{Kim_Jia} in the
superconducting state. As a result, we find \bqa &&
\frac{2\Delta_{c}}{T_{c}} = \mathcal{C}_{BCS} \exp\Bigl( -
\frac{\mathcal{V}_{0}^{-1}}{2N\mathcal{B}_{f}} +
\frac{1}{\sqrt{2N\mathcal{B}_{f}\mathcal{B}_{c}}} \Bigr) , \nn &&
\frac{2\Delta_{f}}{T_{c}} = \mathcal{C}_{BCS} \exp\Bigl( -
\frac{\mathcal{V}_{0}}{\mathcal{B}_{c}} +
\frac{1}{\sqrt{2N\mathcal{B}_{f}\mathcal{B}_{c}}} \Bigr) , \eqa
where $\mathcal{V}_{0} = \frac{\Delta_{c}}{\Delta_{f}}$ is
determined by \bqa && \frac{\mathcal{V}_{0}}{\mathcal{B}_{c}} -
\frac{\mathcal{V}_{0}^{-1}}{2N\mathcal{B}_{f}} = \ln
\mathcal{V}_{0} \eqa and $\mathcal{C}_{BCS} \approx 3.5$ is the
BCS value. Within the range of $T_{c}$ given by Eq. (13), we
obtain $2\Delta_{c}/T_{c} \approx 2.7 \mathcal{C}_{BCS} \sim 9$
while $2\Delta_{f}/T_{c} \sim \mathcal{O}(10^{-1})$.


Recently, thermal conductivity experiments on the heavy fermion
superconductor $CeCoIn_5$ down to 10 $mK$ revealed strong
multi-gap effects with a remarkably low "critical" field for the
small gap band, showing that the complexity of heavy fermion band
structure has a direct impact on their response under magnetic
field \cite{Multi_SC}. In particular, the small gap is claimed to
originate from light electrons instead of heavy fermions, combined
with previous measurements. This aspect seems to be not consistent
with the present description, where such a small gap appears from
pairing correlations of heavy fermions, holons, although the gap
to critical temperature ratio, i.e., $2\Delta_{c}/T_{c} \approx
2.7 \mathcal{C}_{BCS} \sim 9$ and $2\Delta_{f}/T_{c} \sim
\mathcal{O}(10^{-1})$ matches with $CeCoIn_{5}$ \cite{Multi_SC}.
In the Kondo fluctuation mechanism it seems to be natural that the
small gap arises from heavy fermions. We believe that this point
should be clarified in experiments, particulary, from the
measurement for $CeRhIn_{5}$, where the pairing glue in this
superconducting material is claimed to be some local excitations
associated with Kondo fluctuations \cite{TS_Nature}.

\subsection{A fingerprint of the Kondo fluctuation mechanism}

The hallmark of the spin-fluctuation-induced $d-wave$
superconductivity was argued to be emergence of the spin-resonance
mode at an antiferromagnetic wave vector
\cite{Chubukov_Review,Pines_Review}. Since the
hybridization-induced superconductivity allows the $d-wave$
pairing symmetry, the similar spin-resonance mode is expected to
result from pairing correlations of conduction electrons. An
important ingredient beyond the spin-fluctuation scenario is
$d-wave$ pairing of spinless fermions. We claim that emergence of
a charge-resonance mode at an antiferromagnetic wave vector is one
fingerprint of the hybridization-induced superconductivity.

We introduce repulsive interactions between nearest neighbor
holons, given by $H_{int}^{f} = U_{f} \sum_{\langle i j \rangle}
n_{i}^{f} n_{j}^{f}$, where on-site repulsive interactions do not
appear due to the Pauli exclusion principle. Then, the charge
susceptibility is given by the standard RPA
(random-phase-approximation) form \bqa && \chi_{c}^{f}(q,i\Omega)
= \frac{U_{ff}(q)}{1 - U_{ff}(q) \Pi_{c}^{f}(q,i\Omega)} \nonumber
\eqa with $U_{ff} = 2 U_{f} \sum_{j=1}^{d} \cos q_{j}$. It was
shown that $\Im\Pi_{c}^{f}(Q,\Omega < 2 \Delta_{f}) = 0$ and it
jumps at $\Omega = 2 \Delta_{f}$ as $\Im\Pi_{c}^{f}(Q, 2
\Delta_{f} - \epsilon) \not= \Im\Pi_{c}^{f}(Q, 2
\Delta_{f}+\epsilon)$ with $\epsilon \rightarrow 0$, resulting
from $d-wave$ pairing symmetry \cite{Chubukov_Review}, where $Q$
is an associated antiferromagnetic wave vector. The presence of
jump gives rise to the logarithmic singularity in the real part of
the susceptibility as $\Re \Pi_{c}^{f}(Q,\Omega) \propto -
\Delta_{f} \ln \frac{2\Delta_{f}}{|\Omega - 2 \Delta_{f}|}$ via
the Kramers-Kronig relation \cite{INS_Scalapino}. As a result, the
resonance condition of $1 - U_{ff}(Q) \Re
\Pi_{c}^{f}(Q,\Omega_{res}) = 0$ can be always satisfied, causing
a coherent peak in the susceptibility. This is exactly the origin
of the spin-resonance mode in the $d-wave$ superconducting state.
An important point is that holons do not carry spin quantum
numbers but only charge quantum numbers, thus this peak is
identified with a charge-resonance mode at the same momentum with
the spin-resonance mode. This is an essential prediction of the
present mechanism.

\section{Discussion and summary}

\subsection{Comparison with other theoretical frameworks}

An important ingredient in the hybridization-induced mechanism is
the presence of an anomalous propagator of spinon excitations
associated with antiferromagnetic correlations, allowing the
ladder diagram process as the superconducting mechanism (Fig. 2).
One can perform the similar t-matrix calculation at the Kondo
breakdown QCP of the slave-boson theory. Actually, this was
studied in the context of the valance-fluctuation-induced $d-wave$
superconductivity inside the heavy-fermion phase \cite{Miyake}.
Extending this mechanism at the Kondo breakdown QCP, one can
construct particle-particle t-matrices for both conduction
electrons and fermionic spinons. An essential difference from the
slave-fermion theory is that the pairing channel arises from
crossed diagrams instead of ladder diagrams due to the absence of
antiferromagnetic correlations, mathematically corresponding to
the pairing term of bosonic holons in the slave-boson theory (Fig.
2). Since these crossed diagrams involve momentum integrals, such
instability channels become much weaker \cite{Chubukov_Vertex}
than those of the slave-fermion theory.

One can modify the valance-fluctuation mechanism at the Kondo
breakdown QCP, taking into account not only particle-hole pairs
between conduction electrons and fermionic spinons but also their
particle-particle pairs. Recently, this was proposed in the SU(2)
slave-boson formulation of the uniform mean-field ansatz with two
channels for conduction electrons \cite{Coleman}. Another SU(2)
formulation is possible in the $d-wave$ pairing ansatz with one
channel, basically an extended version of the RVB
superconductivity \cite{Senthil_Vojta_Sachdev}. However, these
ideas overestimate quantum fluctuations in spin dynamics, thus
have difficulty in describing antiferromagnetism.

\subsection{Robustness of the $z = 3$ antiferromagnetic quantum criticality
and marginal Fermi liquid phenomenology}

Antiferromagnetism described by the Schwinger boson theory has its
characteristic feature, that is, strong ferromagnetic fluctuations
when "holes" are doped. Physically, such uniform spin fluctuations
result from the fact that the energy dispersion of bosonic spinons
has degeneracy at both the ferromagnetic and antiferromagnetic
wave vectors. This seems to be an important nature of quantum
antiferromagnets, associated with strong quantum fluctuations.
Hybridization fluctuations or "Fermi surface" fluctuations give
rise to Landau damping, resulting in the $z = 3$ antiferromagnetic
QCP. Actually, such strong ferromagnetic fluctuations have been
observed in the YbRh$_{2}$Si$_{2}$-type sample \cite{Ferro}.

If we consider different kinds of orders, an important point is
whether the dispersion of bosonic spinons has its minimum at the
$q = (0, 0, 0)$ momentum point or not. If the energy minimum is
away from $q = (0, 0, 0)$, ferromagnetic fluctuations cannot be
critical and the Landau damping term will not affect critical
spinon dynamics so much because critical spinon excitations appear
in different momentum points which cannot feel such damping. In
the present problem the q = $(0, 0, 0)$ point is almost degenerate
with the q = $(\pi, \pi, \pi)$ because spinons are in an almost
flat band, i.e., $\alpha \ll 1$ in our mathematical expression,
thus allowing strong ferromagnetic fluctuations at the
antiferromagnetic QCP. This is the key physics for the $z = 3$
antiferromagnetic quantum criticality.

Suppose a certain $z = 3$ QCP in three spatial dimensions. Why
does not the marginal Fermi liquid physics arise in such all
systems?

The Fermi surface problem in higher dimensions than one dimension
is extremely difficult. The usually resorted technique so called
large N, where N represents the number of fermion flavors, is not
well defined in the presence of a Fermi surface, basically
originating from bunch of particle-hole soft modes, where all
kinds of planar diagrams, not only self-energy corrections but
also vertex corrections, should be resumed
\cite{FS_Quantum1,FS_Quantum2}, but of course, we do not know how.
In this kind of problems we have two kinds of self-energy
corrections. One is the fermion self-energy while the other is the
boson self-energy. Although we cannot give a definite answer, it
seems that the boson self-energy is determined by the Landau
damping form, given by the self-consistent one-loop calculation,
Eliashberg theory. Actually, this was checked explicitly in the
two-loop order \cite{FS_Quantum1,Z3QCP}. In our opinion this
"protection" mechanism may be due to the presence of a Fermi
surface. Particle-hole excitations around the Fermi surface would
always give rise to the Landau damping around the zero momentum
beyond any order. Ironically, the presence of the Fermi surface
causes a serious problem to self-energy corrections of fermions.
Such calculations in the fermion self-energy require vertex
corrections inevitably \cite{FS_Quantum1}. Actually, this has been
discussed in the community for a long time, but there is still no
consensus on the explicit expression of the fermion Green's
function \cite{Z3QCP}. In this respect the actual exponent for the
fermion self-energy is not known yet.

The main difference from the above problem is that the present
problem consists of two bands, where one is normal but the other
is almost flat. Although it is not completely confirmed, some
arguments are given, associated with the physical reason why
vertex corrections can be neglected in the present two-band model
\cite{Paul,Pepin,Kim_Boltzman}. It is basically due to the fact
that the presence of heavy particles allows us to ignore vertex
corrections because the coefficient $\alpha \ll 1$ appears in the
vertex expression.

In summary, maybe the presence of two bands, more precisely, an
almost flat band allows us to consider only self-energy
corrections, giving rise to the marginal Fermi liquid physics. If
we are dealing with the one-band problem, we should take into
account vertex corrections and we do not know whether the
expression of the fermion Green's function is consistent with the
marginal Fermi liquid form or not. Of course, this discussion is
based on the assumption that the $z = 3$ QCP is stable. Actually,
it turns out that the $z = 3$ quantum criticality is difficult to
be stable if vertex corrections are introduced in the one-band
model \cite{Z3QCP}.

\subsection{Summary}

In this paper we found new mechanism of superconductivity from a
non-Fermi liquid metal beyond the spin fluctuation framework,
originated from strong correlations (Table II). The hybridization
mechanism should be regarded robust and unique, where
antiferromagnetic correlations play an important role in the
presence of hybridization fluctuations at the QCP \cite{Comment},
implying that the similar Kondo mechanism is difficult to work
around the Kondo breakdown QCP in the slave-boson framework. We
predicted emergence of the charge resonance mode at an
antiferromagnetic wave vector as the fingerprint for the Kondo
fluctuation mechanism, resulting from the multi-gap nature, thus
discriminated from the spin fluctuation scenario. We obtain actual
numerical values for the transition temperature and ratio between
the superconducting gaps and transition temperature, and find
$2\Delta_{c} / T_{c} \sim 9$ and $2\Delta_{f} / T_{c} \sim
\mathcal{O}(10^{-1})$. Although these ratios are consistent with
$CeCoIn_{5}$, the origin of each gap is not compatible with an
experiment \cite{Multi_SC}, where the small gap is claimed to
appear from light electrons while it is originated from heavy
fermions, holons in the Kondo fluctuation mechanism. We believe
that this point should be clarified in experiments, particulary,
from the measurement for $CeRhIn_{5}$, where the mechanism of
superconductivity in this material is claimed to differ from that
in $CeCoIn_{5}$ \cite{TS_Nature}.

\begin{table}[ht]
\begin{tabular}{ccccc}
\hline \hline \nn SC from FL & & SC from NFL \nn & Weak coupling &
Strong coupling \nn \hline BCS (Phonon) & Spin-fluctuation &
Kondo-fluctuation \nn mechanism & mechanism & mechanism \nn \hline
\hline
\end{tabular}
\caption{Mechanism of superconductivity around heavy-fermion QCPs
with SC (superconductivity), FL (Fermi liquid), and NFL (non-Fermi
liquid)}
\end{table}

Fruitful discussions with T. Takimoto, J.-H. Han and T. Park are
appreciated. K.-S. Kim thanks B. Fauque for pointing out Ref.
\cite{Multi_SC}. This work was supported by the National Research
Foundation of Korea (NRF) grant funded by the Korea government
(MEST) (No. 2009-0074542).


\begin{thebibliography}{9}
\bibitem{PALee_Review} P. A. Lee, Science {\bf 321}, 1306 (2008).
\bibitem{HF_Review} P. Gegenwart, Q. Si, and F. Steglich,
Nature Physics {\bf 4}, 186 (2008); H. v. Lohneysen, A. Rosch, M.
Vojta, and P. Wolfle, Rev. Mod. Phys. {\bf 79}, 1015 (2007).
\bibitem{BCS} J. R. Schrieffer, \textit{Theory of Superconductivity} (Westview Press,
1999).
\bibitem{MFL_Varma} C. M. Varma, Z. Nussinov, and W. v. Saarloos,
Phys. Rep. {\bf 361}, 267 (2002).
\bibitem{HMM} T. Moriya and J. Kawabata, J. Phys. Soc. Jpn. {\bf
34}, 639 (1973); T. Moriya and J. Kawabata, J. Phys. Soc. Jpn.
{\bf 35}, 669 (1973); J. A. Hertz, Phys. Rev. B {\bf 14}, 1165
(1976); A. J. Millis, Phys. Rev. B {\bf 48}, 7183 (1993).
\bibitem{Chubukov_Review} A. V. Chubukov, D. Pines, and J.
Schmalian, in The Physics of Superconductors, edited by K. H.
Bennemann and J. B. Ketterson (Springer, New York 2003), Vol. 1,
p. 495.
\bibitem{Pines_Review} P. Monthoux, D. Pines, and G. G. Lonzarich,
Nature {\bf 450}, 1177 (2007).
\bibitem{Chubukov_Vertex} A. V. Chubukov, Phys. Rev. B {\bf 72}, 085113
(2005).
\bibitem{GR} R. Kuchler, N. Oeschler, P. Gegenwart,
T. Cichorek, K. Neumaier, O. Tegus, C. Geibel, J. A. Mydosh, F.
Steglich, L. Zhu, and Q. Si, Phys. Rev. Lett. {\bf 91}, 066405
(2003).
\bibitem{TR} J. Custers, P. Gegenwart, H. Wilhelm,
K. Neumaier, Y. Tokiwa, O. Trovarelli, C. Geibel, F. Steglich, C.
Pepin, and P. Coleman, Nature {\bf 424}, 524 (2003).
\bibitem{dHvA} H. Shishido, R. Settai, H. Harima, and Y. Onuki,
J. Phys. Soc. Jpn. {\bf 74}, 1103 (2005).
\bibitem{INS_Local_AF} A. Schroder, G. Aeppli, R. Coldea, M. Adams,
O. Stockert, H.v. Lohneysen, E. Bucher, R. Ramazashvili, and P.
Coleman, Nature {\bf 407}, 351 (2000).
\bibitem{Hall} S. Paschen, T. Luhmann, S. Wirth, P. Gegenwart,
O. Trovarelli, C. Geibel, F. Steglich, P. Coleman, and Q. Si,
Nature {\bf 432}, 881 (2004).
\bibitem{Kim_GR} K.-S. Kim, A. Benlagra, and C. P\'epin,
Phys. Rev. Lett. {\bf 101}, 246403 (2008).
\bibitem{Kim_TR} K.-S. Kim and C. P\'epin,
Phys. Rev. Lett. {\bf 102}, 156404 (2009).
\bibitem{Senthil_Vojta_Sachdev} T. Senthil, S. Sachdev, and M.
Vojta, Phys. Rev. Lett. {\bf 90}, 216403 (2003); T. Senthil, M.
Vojta, and S. Sachdev, Phys. Rev. B {\bf 69}, 035111 (2004).
\bibitem{Paul} I. Paul, C. P\'epin, and M. R. Norman,
Phys. Rev. Lett. {\bf 98}, 026402 (2007); Phys. Rev. B {\bf 78},
035109 (2008).
\bibitem{Pepin} C. P\'epin, Phys. Rev. Lett. {\bf 98}, 206401
(2007); Phys. Rev. B {\bf 77}, 245129 (2008).
\bibitem{OSMT_HF} L. De Leo, M. Civelli, and G. Kotliar,
Phys. Rev. Lett. {\bf 101}, 256404 (2008).
\bibitem{TS_Nature} T. Park, V. A. Sidorov, F. Ronning, J.-X. Zhu,
Y. Tokiwa, H. Lee, E. D. Bauer, R. Movshovich, J. L. Sarrao, and
J. D. Thompson, Nature  {\bf 456}, 366 (2008).
\bibitem{Multi_SC} G. Seyfarth, J. P. Brison, G. Knebel, D. Aoki, G. Lapertot,
and J. Flouquet, Phys. Rev. Lett. {\bf 101}, 046401 (2008).
\bibitem{Coleman_Pepin_Si} P. Coleman, P\'epin, Q. Si, R. Ramazashvili,
Journal of Physics: Condensed Matter {\bf 13}, 723 (2001).
\bibitem{PWAnderson} P. W. Anderson, Nature Physics {\bf 2}, 626
(2006); J. K. Jain and P. W. Anderson, Proc. Natl. Acad. Sci. {\bf
106}, 9131 (2009).
\bibitem{Sachdev_SpN} N. Read and S. Sachdev,
Phys. Rev. Lett. {\bf 66}, 1773 (1991).
\bibitem{Kim_Jia} Ki-Seok Kim and Chenglong Jia, arXiv:0906.0834 (unpublished).
\bibitem{Miyake} Y. Onishi and K. Miyake, J. Phys. Soc. Jpn. {\bf 69}, 3955
(2000); A. T. Holmes, D. Jaccard, and K. Miyake, J. Phys. Soc.
Jpn. {\bf 76}, 051002 (2007).
\bibitem{Coleman} R. Flint, M. Dzero, and P. Coleman,
Nature Physics {\bf 4}, 643 (2008).
\bibitem{Dirre} One cautious person may ask the role of
dangerously irrelevant operators in the field theory above the
upper critical dimension. See the discussion on this subject in A.
Benlagra, K.-S. Kim, and C. P\'epin, arXiv:0902.3630.
(unpublished).
\bibitem{Hewson_Book} A. C. Hewson, \textit{The Kondo Problem
to Heavy Fermions} (Cambridge University Press, New York, 1993).
\bibitem{Confinement_Intro} A. M. Polyakov, \textit{Gauge Fields
and Strings} (Harwood Academic Publishers, 1987); E. Fradkin and
S. H. Shenker, Phys. Rev. D {\bf 19}, 3682 (1979).
\bibitem{Deconfinement_ASL} M. Hermele, T.
Senthil, M. P. A. Fisher, P. A. Lee, N. Nagaosa, and X.-G. Wen,
Phys. Rev. B {\bf 70}, 214437 (2004).
\bibitem{Deconfinement_U1SL_SS} S.-S. Lee, Phys. Rev. B 78, 085129
(2008).
\bibitem{Deconfinement_U1SL_KS} Ki-Seok Kim, Phys. Rev. B 72, 245106 (2005).
\bibitem{Deconfinement_Boson} M. A.
Metlitski, M. Hermele, T. Senthil, and M. P. A. Fisher, Phys. Rev.
B 78, 214418 (2008); T. Senthil, L. Balents, S. Sachdev, A.
Vishwanath, and M. P.A. Fisher, Phys. Rev. B {\bf 70}, 144407
(2004); Ki-Seok Kim, Phys. Rev. B {\bf 72}, 035109 (2005).
\bibitem{NCA_CFT} O.
Parcollet, A. Georges, G. Kotliar, and A. Sengupta, Phys. Rev. B
{\bf 58}, 3794 (1998).
\bibitem{Chubukov_Eliashberg_Theory} A. V. Chubukov and J.
Schmalian, Phys. Rev. B {\bf 72}, 174520 (2005).
\bibitem{INS_Scalapino} N. Bulut and D. J. Scalapino, Phys. Rev. B
{\bf 53}, 5149 (1996).
\bibitem{Comment} Unconventional superconductivity can exist in the
antiferromagnetic phase
owing to spinon condensation. This is exactly dual to the
valance-fluctuation-induced unconventional superconductivity in
the heavy-fermion phase described by holon condensation
\cite{Miyake}. If one tries to extend this hybridization mechanism
around each Kondo breakdown QCP, antiferromagnetic correlations
become essential due to the absence of boson condensation.
\bibitem{Ferro} P. Gegenwart, J. Custers, Y. Tokiwa, C. Geibel, and
F. Steglich, Phys. Rev. Lett. {\bf 94}, 076402 (2005).
\bibitem{FS_Quantum1} Sung-Sik Lee, Phys. Rev. B {\bf 80}, 165102 (2009).
\bibitem{FS_Quantum2} M. A. Metlitski and S. Sachdev, arXiv:1001.1153
(unpublished).
\bibitem{Z3QCP} J. Rech, C. P\'epin, and A. V. Chubukov, Phys. Rev. B
{\bf 74}, 195126 (2006).
\bibitem{Kim_Boltzman} K.-S. Kim and C. P\'epin, J. Phys.: Condens. Matter
{\bf 22}, 025601 (2010).
\end{thebibliography}
\end{document}